\documentclass[prd,draft,eqsecnum,showpacs,aps]{revtex4}
\usepackage{latexsym}
\begin{document}
%\draft
\title{Second-order gravitational effects of local inhomogeneities \\
on CMB anisotropies in nonzero-$\Lambda$ flat cosmological models
}
\author{Kenji Tomita}
%\email{tomita@yukawa.kyoto-u.ac.jp}
\affiliation{Yukawa Institute for Theoretical Physics, 
Kyoto University, Kyoto 606-8502, Japan}
\date{\today}

\begin{abstract}
Nonlinear gravitational effects of large-scale local inhomogeneities 
on Cosmic Microwave Background (CMB) anisotropies are studied, 
based on the relativistic second-order  
theory of perturbations in nonzero-$\Lambda$ flat cosmological models,
which has been analytically derived by the present author, and on the
second-order formula of CMB anisotropies derived by Mollerach and Matarrese. 
In this paper we derive the components of the CMB anisotropy power
spectra in the range of $l = 1 - 22$ which are caused by asymmetric
local inhomogeneities on scales of 300 Mpc. Using our results it is
found that there is a possibility to explain the small
north-south asymmetry of CMB anisotropies which has recently been
observed.
\end{abstract}

\maketitle
%\pacs{Valid PACS appear here.
%{\tt$\backslash$\string pacs\{\}} should always be input,
%even if empty.}

%\narrowtext

%ch1 ----------------------------------------------------------
\section{Introduction}
\label{sec:level1}

In most studies of Cosmic Microwave Background (CMB) anisotropies, the
comparison between observed and theoretical quantities have so far
been done, assuming the linear approximation for cosmological
perturbations. It seems to be successful enough to determine the
cosmological parameters \cite{map,spg,komt}. The present state of our
universe is, however, locally complicated and associated with
nonlinear behavior on various 
scales, and so the observed quantities of CMB anisotropies may include
some small effects caused by large-scale local inhomogeneities through
nonlinear process. Recently it has been reported by Eriksen et
al.\cite{erk1,erk2}, Hansen et al.\cite{hans1,hans2,hans3}, Vielva et
al.\cite{viel} and Park\cite{park} that there is 
a non-trivial north-south asymmetry in various quantities about CMB
anisotropies. These observational results may suggest the existence of
the above effect. 

In this paper we study these nonlinear effects of large-scale local
inhomogeneities on CMB anisotropies, based on the relativistic
second-order theory of cosmological perturbations, which we have
recently derived\cite{tom} and on Mollerach and Matarrese's second-order
formula of CMB anisotropies\cite{cmb}. In Sec. II, we show the
second-order perturbations in 
nonzero-$\Lambda$ flat cosmological models  and the
corresponding CMB anisotropies.  In Sec. III, we derive the expressions
for the second-order power ($\Delta C_l$) of CMB anisotropies,
assuming a dipole form of local inhomogeneities.  
In Sec. IV, we derive numerically the first-order and second-order
anisotropy power spectra, and consider the condition that the local
inhomogeneities can cause the observed asymmetry of the CMB
anisotropy, assuming four simple model types of the radial dependence
in local inhomogeneities. It is found that there is a possibility that
the observed decrease of low multipoles of CMB
anisotropy\cite{olv,cont} also may be explained together with the
above asymmetry. The derivation of main equations in Sec. II and III
are shown in Appendix. Concluding remarks follow in Section V.
 
%ch2 ----------------------------------------------------------
\section{Second-order perturbations and CMB anisotropies}
\label{sec:level2}

First we review the background spacetime and the perturbations which
were derived in the previous paper. The background flat model with
dust matter is expressed as 
\begin{equation}
  \label{eq:m1}
 ds^2 =  g_{\mu\nu} dx^\mu dx^\nu = 
a^2 (\eta) [- d \eta^2 + \delta_{ij} dx^i dx^j],
\end{equation}
where the Greek and Latin letters denote $0,1,2,3$ and $1,2,3$,
respectively, and
$\delta_{ij} (= \delta^i_j = \delta^{ij})$ are the Kronecker
delta. The conformal time $\eta (=x^0)$ is related to the cosmic time
$t$ by $dt = a(\eta) d\eta$. The matter density $\rho$ and the scale
factor $a$ have the relations
\begin{equation}
  \label{eq:m2}
\rho a^2 = 3(a'/a)^2 - \Lambda a^2, \quad {\rm and } \quad 
\rho a^3 = \rho_0,
\end{equation}
where a prime denotes $\partial/\partial \eta$,  $\Lambda$ is the
cosmological constant,  $\rho_0$ is an integration constant and the
units $8\pi G = c =1$ are used.

The first-order and second-order metric perturbations
$\mathop{\delta}_1 g_{\mu\nu} (\equiv 
h_{\mu\nu})$ and $\mathop{\delta}_2 g_{\mu\nu} (\equiv
\ell_{\mu\nu})$, respectively, were derived explicitly by imposing the
synchronous coordinate condition:
\begin{equation}
  \label{eq:m3}  
h_{00} = h_{0i} = 0 \quad {\rm and} \quad \ell_{00} = \ell_{0i} = 0.
\end{equation}
Here we show their expressions only in the growing mode:
\begin{eqnarray}
  \label{eq:m4}
h^j_i &=&  P(\eta) F_{,ij}, \cr
\ell^j_i &=& P(\eta) L^j_i + P^2 (\eta) M^j_i + Q(\eta) N^{|j}_{|i} +
C^j_i, 
\end{eqnarray} 
where $F$ is an arbitrary potential function of spatial coordinates
$x^1, x^2$ and $x^3, \ h^j_i = \delta^{jl}h_{li}, \ N^{|j}_{|i} =
\delta^{jl} N_{|li} = N_{,ij}$ and $P(\eta)$ and $Q(\eta)$ satisfy 
\begin{eqnarray}
  \label{eq:m5}
P'' &+& {2a' \over a} P' -1 = 0, \cr
Q'' &+& {2a' \over a} Q' = P - {5 \over 2} (P')^2.
\end{eqnarray}
The three-dimensional covariant derivative $|i$ are defined in the
space with metric $dl^2 = \delta_{ij} dx^i dx^j$ and their suffices
are raised and lowered by use of $\delta_{ij}$. 
The functions $L^j_i$ and $M^j_i$ are defined by
\begin{eqnarray}
  \label{eq:m6}
L^j_i &=& {1 \over 2}\Bigl[-3 F_{,i} F_{,j} -2 F \cdot F_{,ij} + {1 \over 2}
\delta_{ij} F_{,l} F_{,l}\Bigr], \cr
M^j_i &=& {1 \over 28}\Big\{19F_{,il} F_{,jl} - 12 F_{,ij} \Delta F -
3\delta_{ij} \Bigl[F_{,kl} F_{,kl} -(\Delta F)^2 \Bigr]\Big\}
\end{eqnarray} 
and $N$ is defined by
\begin{equation}
  \label{eq:m7}
\Delta N = {1 \over 28} \Bigl[(\Delta F)^2 - F_{,kl}F_{,kl}\Bigr].
\end{equation} 
The last term $C^l_i$ satisfies the wave equation
\begin{equation}
  \label{eq:m8}
\Box C^j_i = {3 \over 14}(P/a)^2 G^j_i + {1 \over 7}\Bigl[P - {5 \over 2}
(P')^2 \Bigr] \tilde{G}^j_i,
\end{equation} 
where $G^j_i$ and $\tilde{G}^j_i$ are second-order traceless and transverse
functions of spatial coordinates, and the operator $\Box$ is defined by
\begin{equation}
  \label{eq:m9}
\Box \phi \equiv g^{\mu\nu} \phi_{;\mu\nu} = -a^{-2}
\Bigl(\partial^2/\partial\eta^2 + {2a' \over a}\partial/\partial \eta -
\Delta \Bigr) \phi. 
\end{equation} 
So $C^j_i$ represents the second-order gravitational waves caused by
the first-order density perturbations. 

The velocity perturbations $\mathop{\delta}_1 u^\mu$ and
$\mathop{\delta}_2 u^\mu$ vanish, i.e. \ $\mathop{\delta}_1 u^0 =
\mathop{\delta}_1 u^i =0$ and $\mathop{\delta}_2 u^0 =
\mathop{\delta}_2 u^i =0$, and the density perturbations are 
\begin{eqnarray}
  \label{eq:m10}
\mathop{\delta}_1 \rho/\rho &=& {1 \over \rho a^2} \Bigl({a'\over a}P'
-1\Bigr) \Delta F, \cr
\mathop{\delta}_2 \rho/\rho &=& {1 \over 2\rho a^2}\Bigl\{{1 \over
2}(1 - {a' \over a}P') (3F_{,l}F_{,l} + 8F\Delta F) +{1 \over 2}P
[(\Delta F)^2 + F_{,kl}F_{,kl}] \cr
 &+& {1 \over 4}\Bigl[(P')^2 - {2 \over 7}{a'\over a}Q'\Bigr] [(\Delta F)^2 -
F_{,kl}F_{,kl}] - {1 \over 7} {a'\over a}PP' [4F_{,kl}F_{,kl} +
3(\Delta F)^2]  \Big\}.
\end{eqnarray} 

Next let us consider the CMB temperature $T = T^{(0)} (1 + \delta
T/T)$, which $T^{(0)}$ is the background temperature and $\delta T/T
(= \mathop{\delta}_1 T/T + \mathop{\delta}_2 T/T)$ is the
perturbations. The 
present temperature $T^{(0)}_o$ is related to the emitted background
temperature $T^{(0)}_e$ at the decoupling epoch by $T^{(0)}_e = (1+
z_e) T^{(0)}_o$, the temperature perturbation $\tau \equiv (\delta
T/T)_e$ at the decoupling epoch is determined by the physical state
before that epoch , and the present temperature perturbations $(\delta
T/T)_o$ is related to $(\delta T/T)_e$ by the gravitational
perturbations along the light ray from the epoch to the present
epoch. The light ray is described using the background wave vector
$k^\mu \ (\equiv dx^\mu/d \lambda)$, where $\lambda$ is the affine
parameter, and its component is $k^{(0)\mu} = (1, -e^i)$, and the ray
is given by $x^{(0)\mu} = [\lambda, (\lambda_0 - \lambda) e^i]$, where
$e^i$ is the directional unit vector.

The first-order temperature perturbation is 
\begin{equation}
  \label{eq:m11}
\mathop{\delta}_1 T/T =  \tau + {1 \over 2}\int^{\lambda_e}_{\lambda_o}
d\lambda P'(\eta) F_{,ij} e^i e^j.
\end{equation}
Using the relation $dP/d\lambda = P'$ and $dF/d\lambda = - F_{,i}
e^i$, this equation is expressed as
\begin{equation}
  \label{eq:m12}
\mathop{\delta}_1 T/T =  \Theta_1 +\Theta_2
\end{equation}
where 
\begin{eqnarray}
  \label{eq:m13}
\Theta_1 &\equiv& \tau - {1 \over 2}[(P' F_{,i})_e - (P' F_{,i})_o] e^i, \cr
\Theta_2 &\equiv& {1 \over 2}\int^{\lambda_e}_{\lambda_o}
d\lambda P''(\eta) F_{,i} e^i.
\end{eqnarray}
$\Theta_1$ and $\Theta_2$ represent the intrinsic and Sachs-Wolfe
effects, respectively. The latter can be divided into the ordinary
Sachs-Wolfe effect $\Theta_{sac}$ and the Integral Sachs-Wolfe effect
$\Theta_{isw}$, where 
\begin{eqnarray}
  \label{eq:m14}
\Theta_{sac} &\equiv& {1 \over 2} [(P''F)_e - (P''F)_o], \cr
\Theta_{isw} &\equiv& {1 \over 2}\int^{\lambda_e}_{\lambda_o}
d\lambda P'''(\eta) F.
\end{eqnarray}
The second-order temperature perturbation is 
\begin{eqnarray}
  \label{eq:m15}
\mathop{\delta}_2 T/T &=& I_1 (\lambda_e) \Bigl[{1 \over 2} I_1 (\lambda_e)
- \tau\Bigr] - [{A^{(1)}_e}' + \tau_{,i} e^i] \int^{\lambda_e}_{\lambda_o}
d{\lambda} A^{(1)} \cr
&-& \int^{\lambda_e}_{\lambda_o} d{\lambda} \Big\{{1 \over 2} {A^{(2)}}' 
+ A^{(1)}{A^{(1)}}' - {A^{(1)}}'' 
\int^{\lambda}_{\lambda_o} d\bar{\lambda} A^{(1)}(\bar{\lambda})
\Big\} + {\partial\tau \over \partial d^i} d^{(1)i},
\end{eqnarray}
where $(\eta, x^i) = (\lambda, \lambda_o - \lambda)$ in the 
integrands and
\begin{eqnarray}
  \label{eq:m16}
I_1 (\lambda_e) &=& - {1 \over 2} \int^{\lambda_e}_{\lambda_o}
d{\lambda} P' F_{,ij} e^i e^j, \cr
A^{(1)} &=& - {1 \over 2}  P F_{,ij} e^i e^j , \cr
A^{(2)} &=& - {1 \over 2} [P L^j_i + P^2 M^j_i + Q N_{,ij} +
C^j_i ] e^i e^j. 
\end{eqnarray}

Now let us assume an appropriate form of $F$ as follows, to represent a
situation including local inhomogeneities:
\begin{equation}
  \label{eq:m17}
F({\bf x}) = F_P({\bf x}) + F_L({\bf x}),
\end{equation}
where the part of primordial density perturbations ($F_P$) and the
part of local homogeneities ($F_L$) are expressed as
\begin{eqnarray}
  \label{eq:m18}
F_P &=& \int d{\bf k} \alpha ({\bf k}) e^{i{\bf kx}}, \cr
F_L &=& R(r) Y_l^m (\theta, \phi).
\end{eqnarray}
In the former equation, $\alpha({\bf k})$ is a random variable, $Y_l^m
(\theta, \phi)$ is spherical harmonics, and
$(r, \theta, \phi)$ is the polar coordinates, i.e. $r^2 = (x^1)^2 +
(x^2)^2 + (x^3)^2$. For the above $F({\bf x})$, we have the
first-order density perturbation
\begin{eqnarray}
  \label{eq:m19} 
\mathop{\delta}_1 \rho/\rho &=& {1 \over \rho a^2} \Bigl({a'\over a}P'
-1\Bigr) (\Delta F_P + \Delta F_L),\cr
&=& {1 \over \rho a^2} \Bigl({a'\over a}P'-1\Bigr)\Bigl[-\int d{\bf k}
\alpha({\bf k}) k^2 e^{i{\bf kx}} + \tilde{R} (r) Y_l^m (\theta, \phi)\Bigr],  
\end{eqnarray}
where
\begin{equation}
  \label{eq:m20}
\tilde{R} (r) = {1 \over r^2}{d \over dr}(r^2 R_{,r}) - {l(l+1) \over
r^2} R
\end{equation}
with $R_{,r} \equiv d R/dr$.

The part of $\Delta F_P$ is for the ordinary primordial
perturbations, so that by the averaging process, we have
\begin{equation}
  \label{eq:m22}
\langle \alpha ({\bf k}) \alpha ({\bf k'}) \rangle = (2\pi)^{-2}
{\cal P}_F ({\bf k}) \delta ({\bf k} + {\bf k'}),
\end{equation}
with 
\begin{equation}
  \label{eq:m23}
{\cal P}_F ({\bf k}) = {\cal P}_{F0} k^{-3} (k/k_0)^{n-1} T_s^2 (k),
\end{equation}
where $T_s (k)$ is the matter transfer function \cite{sugi} and ${\cal
P}_{F0}$ is the normalization constant. Then the average
value of $(\mathop{\delta}_1 \rho/\rho)^2$ is
\begin{eqnarray}
  \label{eq:m24} 
\langle (\mathop{\delta}_1 \rho/\rho)^2 \rangle &=& \Bigl[{1 \over \rho a^2}
\Bigl({a'\over a}P' -1\Bigr)\Bigr]^2 \Bigl[(2\pi)^{-2} \int d{\bf k} {\cal P}_F (k)
k^4 + (\Delta F_L)^2 \Bigr] \cr
&=& \Bigl[{1 \over \rho a^2} \Bigl({a'\over a}P' -1\Bigr)\Bigr]^2 
\Bigl[(2\pi)^{-2} {\cal P}_{F0}
\int d{\bf k} k (k/k_0)^{n-1} T_s^2(k) + (\Delta F_L)^2 \Bigr]. 
\end{eqnarray}
The part of local inhomogeneities represents a realization of cosmic
variance (in our neighborhood) in primordial perturbations with small $l$.
In order to consider directly the north-south asymmetry, we assume in
the following a simplest case with $l = 1$ and $m = 0$, i.e. $F_L =
R(r) \cos \theta$. As the observational background of this dependence,
there is an asymmetric distribution of galaxies within $200 - 300$ Mpc
around us. According to the studies of galaxy number counts in the Sloan 
Digital Sky Survey (Yasuda et al.\cite{yas}), the galactic number density
in the stripes 
toward the Southern Galactic Cap is larger than that in the stripes
toward Northern Galactic Cap. Such an asymmetry of matter distribution
may extend to the region on these scales. 

Then the first-order temperature perturbations are 
\begin{equation}
  \label{eq:m25}
\mathop{\delta}_1 T/T = \Theta_P + \Theta_L \cos \theta,
\end{equation}
where
\begin{eqnarray}
  \label{eq:m26}
\Theta_P &=& -{1 \over 2} \int d{\bf k} \alpha ({\bf k})
\int^{\lambda_e}_{\lambda_o} d\lambda P' (\eta) (k\mu)^2 e^{i{\bf
kx}}, \cr 
\Theta_L &=& {1 \over 2} \int^{\lambda_e}_{\lambda_o} d\lambda P'
(\eta) R_{,rr},  
\end{eqnarray}
and $\tau$ in Eq.(\ref{eq:m11}) was here neglected, because we pay
attentions to the Sachs-Wolfe effect in the low-$l$ cases such as $l <30$.  
 
The former equation can be rewritten as
\begin{equation}
  \label{eq:m27}
\Theta_P = \int d{\bf k} \alpha ({\bf k}) \Bigl\{- {1 \over 2}[(P'')_o +
ik (P')_o P_1(\mu)] + {1 \over 2} \sum_l (-i)^l (2l+1) {\Theta}_{P(l)}
P_l (\mu) \Bigr\},  
\end{equation}
where 
\begin{equation}
  \label{eq:m28}
{\Theta}_{P(l)} = \int^{\lambda_e}_{\lambda_o} d\lambda  P'''
(\eta) j_l(kr) - \{k(P')_e (2l+1)^{-1}[(l+1) j_{l+1} (kr_e) - l
j_{l-1}(kr_e)] + j_l (kr_e) (P'')_e \}.
\end{equation}
In these equations, we have $\eta = \lambda$ and $r = \lambda_o -
\lambda$. In the derivation of Eq.(\ref{eq:m28}), we used the
relations\cite{zal,hu}  
\begin{equation}
  \label{eq:m29}
e^{i{\bf kx}} = e^{ikr\mu} = \sum_l (-i)^l (2l+1) j_l (kr)
P_l (\mu) 
\end{equation}
and
\begin{equation}
  \label{eq:m30}
(2l +1) \mu P_l(\mu) = (l+1) P_{l+1}(\mu) + l P_{l-1}(\mu),
\end{equation}
where $\mu \equiv \cos  \theta_k$ and $\theta_k$ is the angle between
the vectors $k^i$ and $e^i$. In the above equation for $\Theta_L$,
$\theta$ is the angle between $x^3$-axis and the vector $e^i$,
i.e. $e^1 = \sin \theta \cos \phi, e^2 = \sin \theta \sin \phi, e^3 =
\cos \theta.$

The second-order temperature perturbation consists of three
components:
\begin{equation}
  \label{eq:m31}
\mathop{\delta}_2 T/T = \Theta_{LP} + \Theta_{LL} + \Theta_{PP} ,
\end{equation}
where $\Theta_{LP}$ is a component
including the product of primordial perturbations and local
inhomogeneities, $\Theta_{LL}$ is a component including only
the product of local inhomogeneities, and $\Theta_{PP}$ is a component
including only the product of primordial perturbations. In this paper
we are concerned mainly with studying 
how CMB anisotropies are influenced by local inhomogeneities, and
$\Theta_{PP}$ is small enough, compared with $\Theta_{P}$ in the
first-order. From now, therefore, we neglect $\Theta_{PP}$ and treat
only $\Theta_{LL}$ and $\Theta_{LP}$.

In Eq. (\ref{eq:m15}) for $\mathop{\delta}_2 T/T$, the term with
$A^{(2)}$ includes $P^j_i, M^j_i, N^{|j}_{|i}$ and $C^j_i$. The terms
with $N^{|j}_{|i}$ and $C^j_i$ are small, compared with the terms with
$P^j_i$ and $M^j_i$, because we have $Q/P^2 < 10^{-2}$ always and the
contribution of gravitational radiation is very small. So they are
neglected in the following.

For $\Theta_{LL}$, we obtain from Eq. (\ref{eq:m15})
\begin{equation}
  \label{eq:m32}
\Theta_{LL} = \Theta_{LL}^{(0)} + \Theta_{LL}^{(2)} \cos^2 \theta,
\end{equation}
where
\begin{equation}
  \label{eq:m33}
\Theta_{LL}^{(0)} = {1 \over 16} \int^{\lambda_e}_{\lambda_o} d\lambda
P' (\eta)r^{-2} \Bigl\{R^2 +{2 \over 7}P [13(R_{,r})^2 +12R_{,r}R/r -
6(R/r)^2] \Bigr\},  
\end{equation}
\begin{eqnarray}
  \label{eq:m34}
\Theta_{LL}^{(2)} &=& {1 \over 8}\Bigl(\int^{\lambda_e}_{\lambda_o} d\lambda
P' R_{,rr}\Bigr)^2 - {1 \over 4} \int^{\lambda_e}_{\lambda_o} d\lambda PP'
(R_{,rr})^2 + {1 \over 4} \int^{\lambda_e}_{\lambda_o} d\lambda
P''R_{,rr} \int^{\lambda}_{\lambda_o} d\bar{\lambda} P(\bar{\lambda})
R_{,rr} (\bar{\lambda}) \cr
&+& {1 \over 4}\int^{\lambda_e}_{\lambda_o} d\lambda P' \Bigl\{-{5 \over 4}
(R_{,r})^2 - RR_{,rr} - {1 \over 4} (R/r)^2 \cr
&+& {1 \over 14} P [7(R_{,rr})^2 -7 (R_{,r}/r)^2 -6
r^{-1}R_{,rr}(R_{,r}-R/r^2) -6r^{-3}R(2R_{,r}-R/r)] \Bigr\}. 
\end{eqnarray}

For $\Theta_{LP}$, we obtain from Eq.(\ref{eq:a21}) in Appendix using the
relation (\ref{eq:m30}) and performing partial integrations 
\begin{equation}
  \label{eq:m35}      
\Theta_{LP}/\cos \theta =  {1 \over 4} \int d{\bf k} \alpha ({\bf k}) 
\sum_l (-i)^l (2l+1) {\cal H}_{LP}^{(l)} P_l(\mu), 
\end{equation}
where
\begin{eqnarray}
  \label{eq:m36}
{\cal H}_{LP}^{(l)} &=& \Bigl[- (P'')_e j_l(kr_e) +
\int^{\lambda_e}_{\lambda_o} d\lambda P''' j_l(kr)\Bigr] \times 
\int^{\lambda_e}_{\lambda_o} d\bar{\lambda} P'(\bar{\lambda})
R_{,r}(\bar{\lambda}) \cr
&-& \int^{\lambda_e}_{\lambda_o} d\lambda j_l(kr) \Bigl\{ P'P''R_{,rr} 
+(P''')_e PR_{,rr} +{5 \over 2}{d \over d\lambda} (P' R_{,r}) \cr
&+& P' R_{,rr} + {1 \over 7} {d^2 \over d\lambda^2} [7 P'R
+PP'(4R_{,rr} +9 R_{,r}/r -9 R/r^2)] \cr
&-& {3 \over 7} PP' k^2 (R_{,rr} -R_{,r}/r + R/r^2) 
- {d^2 \over d\lambda^2} \Bigl[P'' \int^{\lambda}_{\lambda_o} d\bar{\lambda}
P(\bar{\lambda}) R_{,rr}(\bar{\lambda})\Bigr]
\Bigr\}  \cr
&+& \int^{\lambda_e}_{\lambda_o} d\lambda P''R_{,rr} 
\int^{\lambda}_{\lambda_o} d\bar{\lambda} P''(\bar{\lambda})
j_l(k\bar{r})
+ \int^{\lambda_e}_{\lambda_o} d\lambda R_{,rr} \{k j_l^{(1)}(kr_e)
[(P')_e P' + (P)_e P'']\cr
&+& k^2 j_l^{(2)}(kr_e) (P')_e P \},
\end{eqnarray}
and we neglected the terms with $\phi_k$ in Eq.(\ref{eq:a21}) because they
vanish in the ${\bf k}$ integration. Since the primordial random
perturbations are rotationally symmetric and the local inhomogeneity
is assumed to 
be proportional to $\cos \theta$, it is reasonable that $\Theta_{LP}$
is represented as the product of $\cos \theta$ and rotationally symmetric
perturbations (cf. Eq.(\ref{eq:m35}). The latter can be expanded using
the Legendre polynomials.

In the above equations, we used the auxiliary functions defined by
\begin{eqnarray}
  \label{eq:m37} 
j_l^{(1)}(kr) &=& {1 \over 2l+1} [l j_{l-1} (kr) - (l+1) j_{l+1}
(kr)], \cr 
j_l^{(2)}(kr) &=& {1 \over 2l+1} \Bigl[{(2l^2+2l-1)(2l +1) \over
(2l-1)(2l+3)} j_l (kr) -
{l(l-1) \over 2l-1} j_{l-2} (kr) - {(l+1)(l+2) \over 2l+3} j_{l+2}
(kr)\Bigr]. 
\end{eqnarray}
%

%ch3 ----------------------------------------------------------
\section{Power spectra of CMB anisotropies}
\label{sec:level3}

The CMB anisotropies in the present analysis include the contributions
from primordial perturbations and local homogeneities, and only the
former perturbations are regarded as statistically random
quantities. In order to derive the power spectra, therefore, we take
the statistical average $\langle \rangle$ only for the contribution
from the primordial perturbations, and $\langle (\delta T/T)^2
\rangle$ is expressed as
\begin{equation}
  \label{eq:c1}
\langle (\delta T/T)^2 \rangle = \langle (\mathop{\delta}_1 T/T)^2
\rangle  + 2 \langle \mathop{\delta}_1 T/T \mathop{\delta}_2 T/T \rangle,
\end{equation}
where we took first two terms and neglected higher order terms. 
For the first-order anisotropies, we have
\begin{equation}
  \label{eq:c2}
\langle (\mathop{\delta}_1 T/T)^2 \rangle = \langle (\Theta_P)^2
\rangle  + (\Theta_L)^2 \cos^2 \theta,
\end{equation}
where $\Theta_L$ is defined in Eq.(\ref{eq:m26}) and
\begin{equation}
  \label{eq:c4}
(T_0)^2 \langle (\Theta_P)^2 \rangle =  \sum_l {2l+1 \over 4\pi} C_l.
\end{equation}
The power spectra $C_l$ are
\begin{equation}
  \label{eq:c5}
C_l =  (T_0)^2 \int dk k^2 {\cal P}_F (k)
|{\cal H}_P^{(l)} (k) |^2, 
\end{equation}
where $T_0$ is the present CMB temperature and
\begin{eqnarray}
  \label{eq:c6} 
{\cal H}_P^{(0)} (k) &=& -(P'')_o - k(P')_e j_1(kr_e) - (P'')_e j_0(kr_e)
+ \int^{\lambda_e}_{\lambda_o} d\lambda P''' j_0(kr), \cr
{\cal H}_P^{(1)} (k) &=& {1 \over 3}k[(P')_o - (P')_e] j_1^{(1)}(kr_e)
-(P'')_e j_1 (kr_e) + \int^{\lambda_e}_{\lambda_o} d\lambda P''' j_1
(kr). 
\end{eqnarray}
For $l \geq 2$, we have
\begin{equation}
  \label{eq:c7}
{\cal H}_P^{(l)} (k) = k (P')_e j_l^{(1)} (kr_e) - (P'')_e j_l (kr_e) +
\int^{\lambda_e}_{\lambda_o} d\lambda P''' j_l (kr). 
\end{equation}
In the derivation of $C_l$, we used Eq.(\ref{eq:m22}) for $\langle
\alpha ({\bf k}) \alpha^* ({\bf k'}) \rangle$, and the formulas for
$P_l (\mu): \ \int^1_{-1} P_l (\mu) P_{l'} (\mu) d\mu = 2/(2l+1), 0$ \
for \ $l=l', l \ne l'$, respectively.

For the second term of Eq. (\ref{eq:c1}), we have
\begin{equation}
  \label{eq:c8}
\langle \mathop{\delta}_1 T/T \mathop{\delta}_2 T/T \rangle = \langle
\Theta_P \Theta_{LP} \rangle + \Theta_L \Theta_{LL} \cos \theta, 
\end{equation}
where
\begin{equation}
  \label{eq:c9}
\Theta_{LL} \cos \theta = \Theta_{LL}^{(0)} \cos \theta + \Theta_{LL}^{(2)} \cos^3 \theta
\end{equation}
and
\begin{equation}
  \label{eq:c10}
\langle \Theta_P \Theta_{LP} \rangle /\cos \theta = 
{1 \over 4} (2\pi)^{-1} \sum_l (2l+1)
\int dk k^2 {\cal P}_F (k) {\cal H}_P^{(l)} {\cal H}_{PL}^{(l)}. 
\end{equation}
If we put $\langle \Theta_P \Theta_{LP} \rangle$ in the form
\begin{equation}
  \label{eq:c11}
(T_0)^2 \langle \Theta_P \Theta_{LP} \rangle =  \sum_l {2l+1
\over 4\pi} \Delta C_l \cos \theta,
\end{equation}
we have 
\begin{equation}
  \label{eq:c12}
 \Delta C_l = {1 \over 2} (T_0)^2 \int dk k^2 {\cal P}_F (k) {\cal
H}_P^{(l)} {\cal H}_{PL}^{(l)},    
\end{equation}
where ${\cal H}_P^{(l)}$ is expressed by performing the $\lambda$
differentiation in Eq.(\ref{eq:m36}) as
\begin{eqnarray}
  \label{eq:c13}
{\cal H}_{PL}^{(l)} &=& \int^{\lambda_e}_{\lambda_o} d\lambda \ \Phi \
j_l(kr) + [- (P'')_e j_l(kr_e) + 
\int^{\lambda_e}_{\lambda_o} d\lambda P''' j_l(kr)] \times 
\int^{\lambda_e}_{\lambda_o} d\bar{\lambda} P'(\bar{\lambda})
R_{,r}(\bar{\lambda}) \cr
&+& \int^{\lambda_e}_{\lambda_o} d\lambda P''R_{,rr} 
\int^{\lambda}_{\lambda_o} d\bar{\lambda} P''(\bar{\lambda})
j_l(k\bar{r})
+ \int^{\lambda_e}_{\lambda_o} d\lambda R_{,rr} \{k j_l^{(1)}(kr_e)
[(P')_e P' + (P)_e P'']\cr
&+& k^2 j_l^{(2)}(kr_e) (P')_e P \},
\end{eqnarray}
where
\begin{eqnarray}
\label{eq:c13b} 
\Phi \equiv &-&PP''R_{,rrr}+ \Bigl[2P'''P + {1 \over 2}P' -(P''')_e
P\Bigr] R_{,rr} - {1 \over 2}P''R_{,r} - P''' R \cr
&-&{1 \over 7}(3P'P''+PP''')(4R_{,rr}+9R_{,r}/r -9R/r^2)+ 
P'''' \int^{\lambda}_{\lambda_o} d\bar{\lambda} PR_{,rr} \cr 
&+&{2 \over 7}[(P')^2+PP''](4R_{,rrr}+9R_{,rr}/r -18R_{,r}/r^2
+18R/r^3) \cr
&-&{1 \over 7}PP'(4R_{,rrrr}+9R_{,rrr}/r -27R_{,rr}/r^2
+54R_{,r}/r^3 -54R/r^4) \cr
&+& {3 \over 7}PP'k^2 (R_{,rr} -R_{,r}/r +R/r^2).
\end{eqnarray}
It is found by numerical calculations that the dominant contribution to
$\Delta C_l$ comes from the last two terms (proportional to $PP'$) in
$\Phi$, especially the terms with highest-order differentiations with
respect to $r$. 

The total anisotropies in Eq.(\ref{eq:c1}) consist of the rotationally
symmetric component ($\langle (\Theta_P)^2 \rangle$), the asymmetric
component 
($\langle \Theta_P \Theta_{LP} \rangle$), and the dipole and
quadrupole components with $(\Theta_L)^2$ and $\Theta_L
\Theta_{LL}$. The key point in this paper is that the north-south
asymmetry affecting the CMB power spectra of $l >2$ is caused by the
component $\langle \Theta_P \Theta_{LP} \rangle$ which is proportional
to $\cos \theta$. 

%ch4 ----------------------------------------------------------
\section{Simple models of large-scale local inhomogeneities}
\label{sec:level4}

In order to study the gravitational influence of local inhomogeneities
on CMB, we consider a simple model with north-south asymmetry, in
which $F_L ({\bf x})$ is expressed as
\begin{equation}
  \label{eq:d1}
F_L ({\bf x}) = R(r) \cos \theta.
\end{equation}
This corresponds to the case $l = 1$ and $m = 0$ in Eq.(\ref{eq:m18}).
For $R$, four types of functional forms are considered and compared :
\begin{eqnarray}
  \label{eq:d2}
R =&& R_0 \exp[-\alpha (x-1)^2], \quad {1 \over 2} R_0 [1 + \cos 2\pi
(x-1)], \cr
&& R_0 x^2 \exp[-\alpha (x-1)^2], \ {\rm and} \ {1 \over 2} R_0 x^2 [1
+ \cos 2\pi (x-1)]
\end{eqnarray}
in the interval $x = [x_1, x_2]$ with $x \equiv r/r_c$, where $x_1
\equiv r_1/r_c = 0.5$ and $x_2 \equiv r_2/r_c = 1.5$. In all types we
have $R = 0$ for $x > x_2$ or $x < x_1$. $R_0$ is the normalization
constant and a constant $\alpha$ is chosen as 20. Here $a_0 r_c$ is assumed to be $\approx 300 h^{-1}$Mpc
($H_0 = 100h$ km/s/Mpc). These types are called in the following
the Gaussian type (G), the sine type (S), the modified Gaussian type
(MG), and the modified sine type (MS), respectively.
In the first two types, R is radially symmetric around the surface $x
= 1$. In the second two types, $R$ has small asymmetry around it.

First, we derive the first-order CMB anisotropies for comparison. They
are calculated using Eq.(\ref{eq:c5}). In Table \ref{table1}, the
behavior of \  $l(l+1) C_l$ \ is shown for $n= 0.97$. The root
mean square $(A_1)$ of \ $l(l+1) C_l \xi$ \ for
$l=2 - 11$ is \ $0.172$ for $n =0.97$, where \ $A_1 \equiv
\{\sum^{11}_{l=2} [l(l+1) C_l]^2/10\}^{1/2} \xi, \ \xi \equiv 2\pi/[{\cal P}_{F0} (T_0)^2]$ and $T_0$ is the present CMB temperature. The
normalization constant ${\cal P}_{F0}$ can be determined
 as ${\cal P}_{F0} = [870 (\mu{\rm K})^2/(2.7{\rm K})^2]
(2\pi)/A_1 = 2.1 \times 10^{-8}$, so that it may be consistent with
the observed CMB anisotropy. 

The first-order anisotropy $\Theta_L$ due to local inhomogeneities is
derived from Eq.(\ref{eq:m26}) in the four types and the ratios of their
values to $R_0$ in the four types are shown in Table \ref{table2}.

Next we derive the second-order anisotropy for local inhomogeneities
of the above four types, using Eq.(\ref{eq:c12}). The behavior of $\Delta
C_l$ is shown in Table \ref{table1} for $l = 1 \sim 22$. It is found
that  $\Delta C_l$ change the signs often in this interval of $l$ ,
but their absolute values are comparable and increase slowly with the
increase of $l$.  The total power is $C_l + 2 \Delta C_l \cos \theta$,
and the north-south asymmetry is represented by the factor $\cos
\theta$ in the second term, which is positive and negative for the
northern and southern hemispheres, respectively.
  
The ratios ($A_2/A_1$) are shown in Table \ref{table3}, where $A_2$ is
defined as the mean 
square root ($A_2$) of $2 l(l+1) \Delta C_l \xi$ in $l = 2 \sim 11$, 
that is, $A_2 \equiv \{\sum^{11}_{l=2} [2l(l+1) \Delta
C_l]^2/10\}^{1/2} \xi$.  The dipole component  
($l =1$) is separately treated from the multiple components ($l \ge 2$).
Here we remark that
we have not specified the value of $a_0 r_c$ yet, and $R_0$ is arbitrary.
From Table \ref{table3}, we find that, as the mean value, we have 
\begin{equation}
  \label{eq:d3}
|R_0| = 7.5 \times 10^{-5} (A_2/A_1).
\end{equation}
%  \bigskip
%
\begin{table}
\caption{CMB anisotropy powers $l(l+1) C_l$ and $l(l+1) \Delta
C_l$ in the case $n = 0.97$. The latter is caused by the 
coupling of cosmological perturbations and local inhomogeneities of
types G, S, MG and MS.\ Here $\xi \equiv 2\pi/[{\cal
P}_{F0} (T_0)^2]$, and ${\cal P}_{F0}$  and $R_0$ are the normalization 
factors.
\label{table1}}
\begin{tabular}{ccrrrrr}
%\tableline
\colrule
&\multicolumn{1}{c}{$l(l+1) C_l \xi$}&\multicolumn{4}{c}{$
10^{-3}\times 2l(l+1) \Delta C_l \xi/R_0 $}\\ 
$l$ & &G\ \ &S\ \ &MG\ \ &MS\ & \ mean \\
%\tableline
\colrule
$1$ & $4.550$ & $-1.07$ & $-2.37$ & $-0.99$ & $-1.73$ \ & $-1.54$\\ 
$2$ & $0.184$ & $-0.54$ & $-0.61$ & $-0.48$ & $-0.55$ \ & $-0.54$\\ 
$3$ & $0.177$ & $ 0.50$ & $-0.13$ & $ 0.68$ & $-0.016$ \ & $ 0.27$\\ 
$4$ & $0.170$ & $-0.66$ & $-0.86$ & $-0.11$ & $-0.22$ \ & $-0.46$\\ 
$5$ & $0.168$ & $ 1.88$ & $ 0.021$ & $ 1.58$ & $ 0.084$ \ & $0.89$\\ 
$6$ & $0.166$ & $ 1.69$ & $ 0.58$ & $ 2.41$ & $ 1.65$ \ & $ 1.58$\\ 
$7$ & $0.167$ & $ 2.26$ & $ 1.03$ & $-0.031$ & $-0.32$ \ & $0.74$\\ 
$8$ & $0.165$ & $ 2.99$& $ 1.26$ & $2.21$ & $ 0.51$ \ & $1.74$\\ 
$9$ & $0.172$ & $-5.32$& $-3.62$ & $-7.52$& $-3.93$ \ & $-5.10$\\ 
$10$ & $0.173$ & $-1.49$ & $-1.08$ & $-3.44$ & $-3.36$ \ & $-2.34$\\ 
$11$ & $0.179$ & $-5.04$ & $ 1.02$ & $-1.69$ & $3.41$ \ & $-0.58$\\ 
$12$ & $0.176$ & $-4.29$ & $-1.79$ & $-3.31$ & $0.43$ \ & $-2.24$\\ 
$13$ & $0.191$ & $5.37$& $ 3.11$& $10.26$& $2.78$ \ & $5.38$\\
$14$ & $0.191$ & $4.26$& $ 2.73$& $6.15$& $7.43$ \ & $5.14$\\
$15$ & $0.203$ & $7.27$ & $-9.40 $ & $4.51$ & $-9.52$ \ & $-1.78$\\ 
$16$ & $0.197$ & $ 2.39$ & $ 3.36 $ & $0.40$ & $-0.84$ \ & $1.33$\\ 
$17$ & $0.215$ & $-2.04$ & $9.66 $ & $-11.56$ & $3.02$ \ & $-0.23$\\ 
$18$ & $0.220$ & $-4.90$ & $-2.12 $ & $-4.33$ & $-4.31$ \ & $-3.92$\\ 
$19$ & $0.230$ & $-19.76$ & $-8.06$ & $-8.44$ & $1.24$ \ & $-8.76$\\ 
$20$ & $0.231$ & $2.80$ & $-6.55$& $3.65$ & $ 2.13$ \ & $0.51$\\  
$21$ & $0.240$ & $32.68$ & $13.26 $ & $35.11$ & $12.92$ \ & $23.49$\\ 
$22$ & $0.260$ & $ 2.49$ & $13.18 $ & $-0.79$ & $1.44$ \ & $4.08$\\ 
%\tableline
\colrule
\end{tabular}
\end{table}
\begin{table}
\caption{CMB anisotropies caused by only local inhomogeneities of
types G, S, MG and MS. $R_0$ is the normalization factor.
\label{table2}}
\begin{tabular}{lccccc}
%\tableline
\colrule
model types &G&S&MG&MS& mean \\
%\tableline
\colrule
$\Theta_L/R_0$\ \ & $0.75$ & $0.022$ & $0.92$ & $0.23$& $0.48$\\
$\Theta_{LL0}/(R_0)^2$\ \ & $-9.1\times 10^3$& $-9.1\times 10^2$ &
$-9.1\times 10^2$ & $-8.7\times 10^2$&$-9.0 \times 10^2$ \\ 
$\Theta_{LL2}/(R_0)^2$\ \ & $2.6\times 10^4$& $1.6\times 10^4$ & $3.1\times
10^4$ & $2.4\times 10^4$& $2.4\times 10^4$\\ 
%\tableline
\colrule
\end{tabular}
\end{table}
\begin{table}
\caption{Ratios of second-order CMB anisotropies ($A_2$) to
first-order CMB anisotropies ($A_1$). The former is caused by the
coupling of cosmological perturbations and local inhomogeneities of
types G, S, MG and MS. $R_0$ is the normalization factor.
\label{table3}}
\begin{tabular}{cccccc}
%\tableline
\colrule
model types &G&S&MG&MS& mean\\
%\tableline
\colrule
$(A_2/A_1)/|R_0|$\ \ &\ $1.6\times 10^4$&\ $0.81\times 10^4$ &\ $1.7\times
10^4$ &\ $1.2\times 10^4$ &\ $1.3\times 10^4$\\ 
%\tableline
\colrule
\end{tabular}
\end{table}

The second-order anisotropies $\Theta_{LL0}/(R_0)^2$ and
$\Theta_{LL2}/(R_0)^2$ are derived using Eqs.(\ref{eq:m33}) and
(\ref{eq:m34}), and their values in four types are shown in Table
\ref{table2}. For the above values of ${\cal P}_{F0}$ and $R_0$, 
$\Theta_{L}, \Theta_{LL0}$ and $\Theta_{LL2}$ are $\approx 3.6\times
10^{-5}(A_2/A_1), -5.1\times 10^{-6}(A_2/A_1)^2$ and $1.2\times
10^{-3}(A_2/A_1)^2$, respectively. 

In order to examine the significance of $R_0$ given in
Eq.(\ref{eq:d3}), we consider here the first-order density
perturbation due to local inhomogeneities. It is expressed as
\begin{equation}
  \label{eq:d4}
(\mathop{\delta}_1 \rho/\rho)_L = {1 \over \rho a^2} \Bigl({a'\over a}P'
-1\Bigr) \tilde{R} \cos \theta,
\end{equation}
where 
\begin{equation}
  \label{eq:d5}
\tilde{R} = (r_c)^{-2} [(x^2 R_{,x})_{,x}/x^2 - 2R/x^2].
\end{equation}
The ratio ($\delta M/M$) of perturbed mass ($\delta M$) to background
mass ($M$) in the interval $x = [x_1, x_2]$ is defined by
\begin{eqnarray}
  \label{eq:d6}
J &\equiv& \int^{x_2}_{x_1} \int^1_0 (\mathop{\delta}_1 \rho/\rho)_L x^2
dx d\mu/\Big\{{1 \over 3}[(x_2)^3 - (x_1)^3] \int ^1_0 d\mu \Big\} \cr
&=& {3 \over 2(\rho a^2)_0 (r_c)^2} \Bigl({a'\over a}P' -1\Bigr)_0
[(x_2)^3 - (x_1)^3]^{-1}
\int^{x_2}_{x_1} [(x^2 R_{,x})_{,x} - 2 R] dx.
\end{eqnarray}
The factor $[(a'/a)P' -1]$ can be regarded as having the
value at present epoch, because the inhomogeneities are at the place
of $z = 0.05 \sim 0.15$, and it is equal to $-0.456$ for the concordant
background model with $\Omega_0 = 0.27$ and $\Lambda_0 =
0.73$. Moreover, $(\rho 
a^2)_0 (r_c)^2 = 3[a r_c/(c/H_0)]^2$, where we have used the unit $c =
8\pi G = 1$ in this paper. Therefore we have
\begin{equation}
  \label{eq:d7}
J = -7.0 ({300 h^{-1}{\rm Mpc} / a_0 r_c}
)^2 \int^{x_2}_{x_1} [(x^2 R_{,x})_{,x} -2R] dx,
\end{equation}
where $x_2 -1 = -(x_1 -1) = 0.5$ and $c/H_0 = 3000 h^{-1}$ Mpc. After
performing the integration in 
Eq.(\ref{eq:d7}), we obtain
\begin{equation}
  \label{eq:d8}
J = (2.4, 14.0, 13.6, 10.3)\  R_0\  ({300h^{-1}{\rm Mpc}/a_0 r_c})^2,
\end{equation}
for types G, S, MG and MS, respectively, and their mean value is
$\bar{J} = 10.1 \ R_0 \ ({300h^{-1}{\rm Mpc} /
a_0 r_c} )^2$. For $R_0$ given in
Eq.(\ref{eq:d3}), $\bar{J}$ leads to
\begin{equation}
  \label{eq:d9}
\bar{J} = 7.6\times 10^{-4} (A_2/A_1)\ ({300h^{-1}{\rm Mpc}/
a_0 r_c} )^2 \times R_0/|R_0|.
\end{equation}

On the other hand, the observed value of $\langle (\delta M/M)^2
\rangle^{1/2}$ is about $1$ on the scale of $8h^{-1}$Mpc and for the
power spectrum $P(k) \propto k^n$, we have $\delta M/M \propto
M^{-(n+3)/6}  \propto r^{-(n+3)/2} = r^{-1.985}$ for $n = 0.97$.
Since we are considering a local inhomogeneity included in the sphere
of radius $2 a_0 r_c$, its scale is regarded as $4 \epsilon a_0 r_c$\
 ($\simeq 1200 \epsilon h^{-1}$Mpc) with $\epsilon \approx 1$, so that
the value of $\delta M/M$ for the inhomogeneity is
\begin{equation}
  \label{eq:d10}
(\delta M/M)_{\rm power} = (8h^{-1}/4 \epsilon a_0 r_c)^{1.985} /b =
4.8\times 10^{-5} (b \epsilon^{1.985})^{-1} ({300h^{-1}{\rm Mpc} / a_0
r_c} )^{1.985}, 
\end{equation}
where $b$ is the biasing factor\cite{sut,lidd}. If we assume $|\bar{J}| =
(\delta M/M)_{\rm power}$, we obtain from Eqs.(\ref{eq:d9}) and
(\ref{eq:d10})
\begin{equation}
  \label{eq:d11}
A_2/A_1 = 0.063 (b \epsilon^{1.985})^{-1} ({300h^{-1}{\rm Mpc} / a_0
r_c})^{0.015}. 
\end{equation}

For this value of $A_2/A_1$, we have
\begin{equation}
  \label{eq:d12}
|\Theta_L| = 2.3\times 10^{-6}/(b \epsilon^{1.985}), \  \Theta_{LL0} =
-2.0\times 10^{-8}/(b \epsilon^{1.985})^2, \ \Theta_{LL2} = 4.8\times
10^{-6}/(b \epsilon^{1.985})^2 \quad {\rm for} 
\ n=0.97,
\end{equation}
neglecting the factor of $a_0 r_c$.
If $b \epsilon^{1.985} \sim 1$, these are small, compared with
$\sqrt{C_1}/T_0 (\sim 
5\times 10^{-5})$ and $\sqrt{C_2}/T_0 (\sim  10^{-5})$. If  $b
\epsilon^{1.985} < 0.5$, however, $\Theta_{LL2}$ is comparable with  
$\sqrt{C_2}/T_0$ or larger than it.  

Finally let us compare our above theoretical results with the observed
anisotropy spectra by Eriksen et al.\cite{erk1}. Here we assume that
the matter density perturbations in the Southern and Northern
hemispheres are positive and negative, respectively, corresponding to
the galactic number counts in SDSS\cite{yas}. Then we have $R_0 <0$ in
order that the directions $\theta = 0$ and $\pi$ may be in the Galactic
North and South, respectively. In our results of calculations (from
Table \ref{table1}), we find the trend that $\Delta C_l$ for $l = 2, 3
$ and $4$ are positive for $R_0 <0$ and $\Delta C_l$ for $l \ge 5$
change their signs 
in the period of $\Delta l = 3$. This trend seems to be seen in Fig.2
of Eriksen et al.'s paper\cite{erk1}. It is found, moreover, that
corresponding to the north-south asymmetry (proportional to $\cos
\theta$) the ratio $\Delta C_l/C_l$ seems  
to be $0.3 \sim 0.5$ by reading the points in Fig. 2 of their paper.
From Eq.(\ref{eq:d11}) and Eriksen et al.'s observational result,
accordingly, we obtain the condition $b \epsilon^{1.985} = 0.1
\sim 0.2$. This condition can be satisfied in the reasonable range of
$b$. For instance we have $\epsilon = 0.45 \sim 0.63$ for $b =0.5$.

Under this condition, $\Theta_{LL2}$ is comparable with $\sqrt{C_2}/T_0$,
so that the measured value of $C_2$ may be disturbed by
$\Theta_{LL2}$, because of their similar angular dependence.
$\Theta_{LL2}$ and $\sqrt{C_2}$ depend on $P_2(\cos \theta)$ and 
$P_2(\cos \theta_k)$, where $\theta$ and $\theta_k$ are the angles
between a directional vector ${\bf e}$ and the $x^3$-axis and between
${\bf e}$ and ${\bf k}$, respectively (cf. Appendix), but both of them
may be measured as the quadrupole components. 
So its measured value may have been given a value
smaller than its theoretical expected value by the offset effect.   

%ch5 ----------------------------------------------------------
\section{Concluding remarks}
\label{sec:level5}

In this paper we derived the asymmetry of CMB anisotropy powers
(proportional to $\cos \theta$) with comparatively low $l$ using the
second-order 
perturbation theory by assuming the local matter distribution is
dipole-like. This assumption seems to be rough, but it is consistent
with the observed situation that the asymmetry of CMB anisotropies
disappears when they are averaged in the whole sky and the asymmetry
in the distribution of galaxies is also disappears similarly in the
whole sky, though it has only a small north-south asymmetry.
If we add a correction term $\propto \cos^2 \theta$ to $F_L$, more
realistic simulations to observed $\Delta C_l$ may be 
obtained, especially for $l > 20$. Moreover, to clarify the relation 
between the derived asymmetry and non-Gaussianity, it is necessary 
to investigate the multi-point correlations of anisotropies. Detailed 
analyses on these points are to be done in the next step.

In our result, $\Delta C_l/C_l$ depend not so on the distance to the
local inhomogeneity (i.e. the value of $a_0 r_c$), but sensitively on
$b \epsilon^{(n+3)/2}$.  It was found that the behavior of 
$\Delta C_l$ in the case $n = 0.90$ is not so different from that in
the case $n = 0.97$, so the conclusion in our result does not depend
on $n$.  

The northern and southern hemispheres were assumed here as those in the
Galactic coordinate frame. However our theory itself does not depend
on any coordinate frames and can treat any local inhomogeneities which
can be supposed.
  
When the measurements of CMB anisotropies are more accurate, the
relation between the local matter distribution and the CMB asymmetry
will be more realistic through the second-order perturbation theory.

%%%%%%%%%%%%%%%%%%%%%%%%%%%%%%%%%%%%%%%%%%%%%%%%%%%%%%%%%%%%%%%%%%%%
\begin{acknowledgments}
The author thanks K.T. Inoue for helpful discussions on the asymmetry of
CMB anisotropies. Numerical computation in this work was carried out
at the Yukawa Institute Computer Facility.  
\end{acknowledgments}
%%%%%%%%%%%%%%%%%%%%%%%%%%%%%%%%%%%%%%%%%%%%%%%%%%%%%%%%%%%%%%%%%%%%

% appA
%----------------------------------------------
\appendix
\section{Derivation of $\mathop{\delta}_2 T/T$}

From Eq.(\ref{eq:m15}), we obtain
\begin{eqnarray}
  \label{eq:a1}
\mathop{\delta}_2 T/T &=& {1 \over 2}[I_1 (\lambda_e)]^2 
- {A^{(1)}_e}'  \int^{\lambda_e}_{\lambda_o} d{\lambda} A^{(1)} \cr
&-& \int^{\lambda_e}_{\lambda_o} d{\lambda} \Big\{{1 \over 2} {A^{(2)}}' 
+ A^{(1)}{A^{(1)}}' - {A^{(1)}}'' 
\int^{\lambda}_{\lambda_o} d\bar{\lambda} A^{(1)}(\bar{\lambda})
\Big\}, 
\end{eqnarray}
neglecting the terms with $\tau$ at the decoupling epoch. Each term in
Eq.(\ref{eq:a1}) is expressed as follows:
\begin{equation}
  \label{eq:a2}
I_1 (\lambda_e) = -{1 \over 2}\int^{\lambda_e}_{\lambda_o} d{\lambda}
P' (F_{L,ij} e^ie^j + F_{P,ij} e^ie^j), 
\end{equation}
so that
\begin{equation}
  \label{eq:a3}
[I_1 (\lambda_e)]^2 = {1 \over 4}\Bigl(\int^{\lambda_e}_{\lambda_o} d{\lambda}
P' F_{L,ij} e^ie^j\Bigr)^2  + {1 \over 2} \int^{\lambda_e}_{\lambda_o} d{\lambda}
P' F_{L,ij} e^ie^j \int^{\lambda_e}_{\lambda_o} d{\lambda}
P' F_{P,ij} e^ie^j, 
\end{equation}
where the term $(\int^{\lambda_e}_{\lambda_o} d{\lambda} P' F_{,ij}
e^ie^j)$ \ is neglected by the reason described in the text. Similarly we
obtain 
\begin{equation}
  \label{eq:a4}
A^{(1)} {A^{(1)}}' = {1 \over 4}PP' [(F_{L,ij} e^ie^j)^2 + 2F_{L,ij}
e^ie^j F_{P,kl} e^k e^l],
\end{equation}
\begin{equation}
  \label{eq:a5}
{A^{(1)}}'' \int^{\lambda_e}_{\lambda_o} d{\lambda} A^{(1)} = {1 \over
4}P'' \Bigl[F_{L,kl} e^ke^l \int^{\lambda_e}_{\lambda_o} d{\lambda} P
(F_{L,ij} 
+ F_{P,ij})e^ie^j + F_{P,kl} e^k e^l \int^{\lambda_e}_{\lambda_o}
d{\lambda} P F_{L,ij} e^ie^j\Bigr],
\end{equation}
and 
\begin{equation}
  \label{eq:a6}
\int^{\lambda_e}_{\lambda_o} d{\lambda} {A^{(2)}}' = -{1 \over 2}
\int^{\lambda_e}_{\lambda_o} d{\lambda} [P' L^j_i + 2PP' M^j_i]e^ie_j,
\end{equation}
where
\begin{eqnarray}
  \label{eq:a7} 
L^j_i e^ie_j &=& -{3 \over 2} (F_{L,i}e^i)^2 - F_L F_{L,ij}e^ie^j + {1
\over 4} F_{L,l}F_{L,l} - 3F_{L,i}e^iF_{P,j}e^j \cr
&-& (F_L F_{P,ij} + F_P F_{L,ij}) e^ie^j + {1 \over 2}F_{L,l}F_{P,l}, \cr
M^j_i e^ie_j &=& {1\over 28}\{[19 F_{L,il}F_{L,jl} - 12 \Delta F_L
F_{L,ij}] e^ie^j - 3 [F_{L,kl}F_{L,kl} - (\Delta F_L)^2] \cr
&+& [38 F_{L,il}F_{P,jl} - 12 \Delta F_L F_{P,ij}- 12 \Delta F_P 
F_{L,ij}] e^ie^j 
- 6 [F_{L,kl}F_{P,kl} - \Delta F_L \Delta F_P]\}.
\end{eqnarray}
Using the radial coordinate $r$, we have 
\begin{eqnarray}
  \label{eq:a71}
F_{L,j} e^j &=& R_{,r} Y^m_l, \cr
F_{P,j} e^j &=& \int d{\bf k} \alpha({\bf k}) ik \mu e^{i{\bf kx}},  
\end{eqnarray}
corresponding to Eq.(\ref{eq:m18}), where ${\bf kx} = kr\mu$, and 
\begin{eqnarray}
  \label{eq:a72} 
F_{P,j} F_{P,j} &=& i\int d{\bf k} \alpha({\bf k}) F_{L,j} k^j
e^{i{\bf kx}}, \cr
F_{P,jl} F_{P,jl} &=& -\int d{\bf k} \alpha({\bf k}) F_{L,jl} k^j k^l
e^{i{\bf kx}}.
\end{eqnarray}
To express $F_{L,j} k^j$ and $F_{L,jl} k^j k^l$ in terms of polar
coordinates, we introduce
two unit (three-dimensional) vectors $e_\theta^i$
and $e_\phi^i$ together with $e^i$, and make an orthonormal triad
vectors with components: 
\begin{eqnarray}
  \label{eq:a7b} 
(e^i) &=& (\sin \theta \cos \phi, \sin \theta \sin \phi, \cos \theta), \cr 
(e_\theta^i) &=& (\cos \theta \cos \phi, \cos \theta \sin \phi, -\sin
\theta), \cr 
(e_\phi^i) &=& (- \sin \phi, \cos \phi, 0),
\end{eqnarray}
respectively, where $\theta$ is the angle between the vector $e^i$ and
the $x^3$-axis. Using these vectors, we have $\partial r/\partial x^i =
e_i, \ \partial \theta/\partial x^i = r^{-1} e_{\theta i}$ and
$\partial \phi/\partial x^i = (r\sin \theta)^{-1} e_{\phi i}$ for the
polar coordinates $(r, \theta, \phi)$.

Moreover, the projection of ${\bf k}$ to the triad is expressed using
another angles $\theta_k$ and $\phi_k$ as
\begin{equation}
  \label{eq:a8}
k^i (e_{\theta i}, e_{\phi i}, e_i) = k (\sin \theta_k \cos \phi_k,
\sin \theta_k \sin \phi_k, \cos \theta_k),  
\end{equation}
where $e_i = \delta_{ij} e^j$, etc, and the angle $\theta_k$ is the
angle between the vectors $e^i$ and $k^i$ . Then we have
\begin{eqnarray}
  \label{eq:a9}
F_{L,i} k^i &=& F_{L,l} (e^l k^i e_i + e^l_\theta k^i e_{\theta i} +
e^l_\phi k^i e_{\phi i}) \cr
&=& k [F_{L,r} \cos \theta_k + F_{L,\theta}
r^{-1} \sin \theta_k \cos \phi_k + F_{L,\phi} (r\sin \theta)^{-1}
\sin \theta_k \sin \phi_k],
\end{eqnarray}
\begin{eqnarray}
  \label{eq:a10} 
F_{L,ij} k^i k^j &=& k^i k^j [F_{L,rr} e_i e_j + (F_{L,\theta\theta}
+rF_{L,r})r^{-2} e_{\theta i} e_{\theta j} + (F_{L,\phi\phi} +
r\sin^2\theta F_{L,r}  \cr
&+&\sin \theta \cos \theta F_{L,\theta})(r\sin
\theta)^{-2} e_{\phi i} e_{\phi j} \cr
&+& 2(F_{L,r\theta} -r^{-1} F_{L,\theta}) r^{-1} e_i e_{\theta j} +
2(F_{L,\phi\phi} - r^{-1} F_{L,\phi}) (r\sin \theta)^{-1}e_i 
e_{\phi j} \cr
&+& (F_{L,\theta\phi} - \cot \theta F_{L,\phi}) (r^2 \sin \theta)^{-1}
e_{\theta i} e_{\phi j}]. 
\end{eqnarray}

Substituting $F_P$ and $F_L$ in Eq.(\ref{eq:m18}) with $l = 1$ and $m
= 0$ into Eq.(\ref{eq:a3}), we obtain 
\begin{equation}
  \label{eq:a11}
(I_1)^2 = {1 \over 4} \Bigl(\int^{\lambda_e}_{\lambda_o} d{\lambda}
P'R_{,rr}\Bigr)^2 \cos^2 \theta + {1 \over 2}\int^{\lambda_e}_{\lambda_o}
d{\lambda} P' R_{,rr} \cos \theta \int^{\lambda_e}_{\lambda_o}
d{\lambda} P' i \int d{\bf k} \alpha({\bf k}) k\mu e^{i{\bf kx}},
\end{equation} 
\begin{equation}
  \label{eq:a13}
A^{(1)}{A^{(1)}}' = {1 \over 4} PP' \Bigl[(R_{,rr})^2 \cos^2 \theta -
2i R_{,rr} \cos \theta  \int d{\bf k} \alpha({\bf k}) (k\mu)^2 e^{i{\bf
kx}}\Bigr], 
\end{equation} 
\begin{eqnarray}
  \label{eq:a14}
{A^{(1)}}'' \int^{\lambda}_{\lambda_o} d{\lambda} A^{(1)} &=& {1
\over 4} P'' \Bigl[R_{,rr} \int^{\lambda}_{\lambda_o} d{\bar{\lambda}}
P(\bar{\eta}) R_{,rr}(\bar{r}) \cos^2 \theta - R_{,rr}\cos \theta  
\int^{\lambda}_{\lambda_o} d\bar{\lambda} P \int d{\bf k} \alpha({\bf
k}) (k\mu)^2  e^{i{\bf kx}}\cr
&-& \int d{\bf k} \alpha({\bf k}) (k\mu)^2 e^{i{\bf kx}}      
\int^{\lambda}_{\lambda_o} d\bar{\lambda} P R_{,rr}\cos \theta \Bigr]. 
\end{eqnarray}

For the terms in ${A^{(2)}}'$, we have
\begin{equation}
  \label{eq:a15}
F F_{,ij} e^i e^j = RR_{,rr} \cos^2 \theta + \int d{\bf k} \alpha({\bf
k}) e^{i{\bf kx}} [R_{,rr} - (k\mu)^2 R] \cos \theta,  
\end{equation} 
\begin{equation}
  \label{eq:a16}
F_{,l}F_{,l} = (R_{,r})^2 \cos^2 \theta + R^2 \sin^2 \theta/r^2 +
2\int d{\bf k} \alpha({\bf k}) e^{i{\bf kx}} ik [R_{,r}\cos \theta \mu
- r^{-1}R \sin \theta \sin \theta_k \cos \phi_k], 
\end{equation} 
\begin{eqnarray}
  \label{eq:a17}
F_{,il} F_{,jl} e^i e^j &=& (R_{,rr})^2 \cos^2 \theta +  (R_{,r})^2
\sin^2 \theta/r^2 + 2 \int d{\bf k} \alpha({\bf k}) e^{i{\bf kx}}
[-R_{,rr}\cos \theta (k\mu)^2 \cr
&-& R_{,r} \sin \theta k^2\mu \sin \theta_k
\cos \phi_k],  
\end{eqnarray} 
\begin{eqnarray} 
  \label{eq:a18}
\Delta F F_{,ij} e^i e^j &=& (R_{,rr} + 2r^{-1}R_{,r} -2r^{-2}R) R_{,rr}
\cos^2 \theta  \cr
&-&\int d{\bf k} \alpha({\bf k}) e^{i{\bf kx}} [(R_{,rr}
+2r^{-1}R_{,r} -2r^{-2}R)(k\mu)^2 + R_{,rr} k^2] \cos \theta,
\end{eqnarray} 
\begin{eqnarray}
  \label{eq:a19}
F_{,ij} F_{,ij} - (\Delta F)^2 &=& -4r^{-1}(R_{,r} -r^{-1}R) (R_{,rr}
+r^{-1}R_{,r} - 
r^{-2}R) \cos^2 \theta  +2r^{-2}(R_{,r} - r^{-1}R)^2 \cr
&-& 2\int d{\bf k} \alpha({\bf k}) k^2 e^{i{\bf kx}} \{R_{,rr} \mu^2
\cos \theta +r^{-1}(R_{,r} -r^{-1}R)[\cos \theta \sin^2 \theta_k \cr
&-& 2\sin \theta \sin \theta_k \cos \theta_k \cos \phi_k] - (R_{,rr} +
2r^{-1}R_{,r} -2r^{-2}R) \cos \theta \}.
\end{eqnarray} 

Using the above equations, we obtain $\Theta_{LL}$ and $\Theta_{LP}$ as
follows:
\begin{eqnarray}
  \label{eq:a20}
\Theta_{LL} &=& {1 \over 8}\Bigl(\int^{\lambda_e}_{\lambda_o}
d{\lambda} P' R_{,rr}\cos \theta\Bigr)^2 - {1 \over 4}
\int^{\lambda_e}_{\lambda_o} d{\lambda} PP'(R_{,rr}\cos \theta)^2 \cr
&+& {1 \over 4}
\int^{\lambda_e}_{\lambda_o}
d{\lambda} P''R_{,rr}  \int^{\lambda}_{\lambda_o} d{\bar{\lambda}}
P(\bar{\eta}) R_{,rr} (\bar{r})\cos^2 \theta \cr
&+& {1 \over 4} \int^{\lambda_e}_{\lambda_o}
d{\lambda} P'\Bigl\{-\Bigl[RR_{,rr} +{5 \over 4}(R_{,r})^2\Bigr] 
\cos^2 \theta + {1 \over 4} r^{-2}R^2 \sin^2 \theta \cr
&+& {1 \over 14} P[19(R_{,rr})^2 \cos^2
\theta + 19 (R'/r)^2 \sin^2 \theta -12 R_{,rr}(R_{,rr}
+2r^{-1}R_{,r}-2r^{-2}R)\cos^2 \theta \cr
&+& 12r^{-1}(R_{,r}- r^{-1}R)(R_{,rr} +r^{-1}R_{,r} -
r^{-2}R) \cos^2 \theta -6(R_{,r} -r^{-1}R)^2] \Bigr\}
\end{eqnarray} 
and 
\begin{eqnarray}
  \label{eq:a21}
\Theta_{LP} &=& -{1 \over 4}\int^{\lambda_e}_{\lambda_o} d{\lambda} P'
R_{,rr}\cos \theta \times \int^{\lambda_e}_{\lambda_o} d{\lambda} P'
\int d{\bf k} \alpha({\bf k}) (k\mu)^2 e^{i{\bf kx}}   \cr
&+& {1 \over 2}\int^{\lambda_e}_{\lambda_o} d{\lambda} PP'
R_{,rr}\cos \theta \int d{\bf k} \alpha({\bf k}) (k\mu)^2 e^{i{\bf
kx}} \cr
&-& {1 \over 4}\int^{\lambda_e}_{\lambda_o} d{\lambda} P'' \int d{\bf k}
\alpha({\bf k}) (k\mu)^2 e^{i{\bf kx}} \Bigl[R_{,rr}
\int^{\lambda}_{\lambda_o} d{\bar{\lambda}} P(\bar{\eta}) e^{i{\bf
k\bar{x}}} + e^{i{\bf kx}}\int^{\lambda}_{\lambda_o} d{\bar{\lambda}}
P(\bar{\eta})R_{,rr}P(\bar{\eta}) \Bigr] \cos \theta \cr
&+& {1 \over 4}\int^{\lambda_e}_{\lambda_o} d{\lambda} P'\Bigl\{- {5 \over
2}R_{,r} \cos \theta \ i\int d{\bf k} \alpha({\bf k}) k\mu e^{i{\bf
kx}} - \int d{\bf k} \alpha({\bf k}) e^{i{\bf kx}} [R_{,rr} -(k\mu)^2
R] \cos \theta \cr
&-& {1 \over 2} \int d{\bf k} \alpha({\bf k}) e^{i{\bf
kx}} r^{-1}R\sin \theta \ ik \sin \theta_k \cos \phi_k \Bigr\} \cr
&+& {1 \over 28}\int^{\lambda_e}_{\lambda_o} d{\lambda} PP' \int
d{\bf k} \alpha({\bf k}) e^{i{\bf kx}} \{19[-R_{,rr} \cos \theta
(k\mu)^2 + R_{,r}\sin \theta k^2 \mu \sin \theta_k \cos \phi_k] \cr
&+& 6 k^2[(R_{,rr} + 2r^{-1}R_{,r}-2r^{-2}R)\mu^2 +R_{,rr}] \cos \theta 
+ 3k^2[R_{,rr} \cos \theta \mu^2 \cr
&+&r^{-1}(R_{,r}- r^{-1}R)(\cos \theta \sin^2
\theta_k - 2\sin \theta  \sin \theta_k \cos \theta_k \cos \phi_k)
-(R_{,rr} +2r^{-1}R_{,r}-2r^{-2}R) \cos \theta]\} \cr
&+& {1 \over 4}(P')_e \int d{\bf k} \alpha({\bf k}) (k\mu)^2 e^{i{\bf kx}}
\int^{\lambda_e}_{\lambda_o} d{\lambda} PR_{,rr} \cos \theta. 
\end{eqnarray} 

By integrating Eq.(\ref{eq:a21}) partially with respect to $\lambda$,
we obtain Eq.(\ref{eq:m35}). In the process of integrations, it is to
be noticed that we use the boundary condition $(R)_e = (R)_o = 0$ and
$(d^m R/dr^m)_e = (d^m R/dr^m)_o = 0$ for $1 \le m \le 4$.
%
%\begin{eqnarray}
%  \label{eq:a22}

%\end{eqnarray} 
%

% 
%%%%%%%%%%%%%%%%%%%%%%%%%%%%%%%%%%%%%%%%%%%%%%%%%%%%%%%%%%%%%%%%%%%%

%ref
%\begin{references}

%*****************************************************

\end{document}